\documentclass[letterpaper]{jsara} \journalinfo{Journal of the Southeastern
Association for Research in Astronomy, {\bf 6}, 00-00, \today}

\usepackage{pdflscape}

\slugcomment{}

\shorttitle{SX Phoenicis in  NGC 4833}
\shortauthors{Darragh \& Murphy}

\begin{document}

\setcounter{page}{01}


\title{New SX Phoenicis Variables in the Globular Cluster NGC 4833\altaffilmark{1}}


\author{Andrew N. Darragh and Brian W. Murphy}
\affil{Department of Physics and Astronomy, Butler University, 
Indianapolis, IN, 46208 \& SARA}

\altaffiltext{1}{Based on observations obtained with the SARA
  Observatory 0.6 m telescope at Cerro Tololo, which is owned and
  operated by the Southeastern Association for Research in Astronomy
  (http://www.saraobservatory.org). }

\email{bmurphy@butler.edu}

\begin{abstract}
 We report the discovery of 6 SX Phoenicis stars in the southern
 globular cluster NGC 4833. Images were obtained from January through
 June 2011 with the Southeastern Association for Research in Astronomy
 0.6 meter telescope located at Cerro Tololo Interamerican
 Observatory. The image subtraction method of Alard \& Lupton (1998)
 was used to search for variable stars in the cluster. We confirmed 17
 previously cataloged variables by Demers \& Wehlau(1977). In
 addition to the previously known variables we have identified 10 new
 variables. Of the total number of confirmed variables in our 10$\times$10
 arcmin$^2$ field, we classified 10 RRab variables, with a mean period
 of 0.69591 days, 7 RRc, with a mean period of 0.39555 days, 2
 possible RRe variables with a mean period of 0.30950 days, a W Ursae
 Majoris contact binary, an Algol-type binary, and the 6 SX Phoenicis
 stars with a mean period of 0.05847 days. The periods, relative
 numbers of RRab and RRc variables, and Bailey diagram are indicative
 of the cluster being of the Oosterhoff type II. We present the
 phased-light curves, periods of previously known variables and the
 periods and classifications of the newly discovered variables, and
 their location on the color-magnitude diagram.
\end{abstract}




\section{Introduction}

NGC~4833 is a southern globular cluster lying near the Galactic plane
in the constellation Musca. Its variable stars have not been studied
in detail for nearly 35 years since observations by Demers \& Wehlau (1977). The first
survey of variable stars in the cluster's vicinity was done by Bailey
(1924) who found two variables, both of which are likely not cluster
members. Later studies revealed a total of 16 variables in the
vicinity of the cluster (Wright 1941; Sawyer-Hogg 1973).  Among these
were the two variables found by Bailey with most of the remaining
being cluster variables, known today as RR Lyrae variables. Using
photographic plates Demers \& Wehlau (1977) did B and V photometry of
the cluster identifying 24 variables and determined accurate periods
for 15 RR Lyrae variables.  They concluded that the cluster was
Oosterhoff Type II (Oosterhoff 1939).  The most recent study was Melbourne et
al. (2000).  They investigated extinction near NGC~4833 and determined
the positions of the RR Lyrae variables on a color-magnitude diagram
but did not do time-series photometry.  NGC~4833 is 6.6 Kpc from the
Sun, has a core radius of 1.0{\arcmin}, and a half light radius of
2.41{\arcmin} which make it a relatively easy target for searching for
and investigating its variable stars (Harris 1996).

Given the lack of recent detailed study of the variables in NGC~4833 we
added the cluster to our continuing program to study in detail a
number of southern globular star clusters (Conroy et al. 2012, Toddy et al. 2012).  With the combination of
CCD images, an observation window spanning more than 4 months, and an
image subtraction package developed by Alard (2000), we are able to
detect, analyze variable stars, and produce unprecedented high-quality
light curves and periods for each variable.  In this study we present
the preliminary results of our observations of NGC~4833 including a
color-magnitude diagram, as well as a Bailey diagram, complete light
curves, and periods for 27 variables, with 10 of them being newly
identified.

\section{Observations and Analysis}

Images of NGC 4833 were obtained during 14 nights between 30 January
and 25 June, 2011, using the Southeastern Association for Research in
Astronomy (SARA) 0.6 meter telescope located at Cerro Tololo
Interamerican Observatory. In all, nearly 1700 images were obtained in
the V-band.  Also, multicolor sequences were done on 9 March, 1, 3,
and 6 May, 2011, as a VRI sequence on 9 March, and a BVRI sequence on
1, 3, and 6 May.  An Apogee Alta E6 camera was used for the
observations with a 1024$\times$1024 pixel Kodak KAF1001E chip, a gain
of 1.5 electron per ADU count, and an RMS noise of 8.9 electrons. We
used 1$\times$1 binning resulting in an image scale of
0.61{\arcsec}/px resulting in a 10$\times$10 arcmin$^2$ field of
view. On all nights of observation, 120 second exposures were taken
using a Bessel V filter.  The exposure time allowed us to detect
dimmer variables but caused many of the brighter stars in the field
to be overexposed.  Typical seeing ranged between 1.2 and
1.7{\arcsec} in the V-band.

All images were debiased, flat-fielded, dark subtracted, and cleaned
of hot, cold, and bad pixels. Images were then analyzed using the
ISIS2.2 package (Alard \& Lupton 1998, Alard 2000). The package works
by convolving the reference image shown  in Figure 1, composed of a combination of the
highest quality images obtained, to the point spread function of each
individual image, then subtracting each image from the reference image
to determine if any change in intensity has occurred.  A resulting 
variable image identifies possible variables and is shown in Figure 2. A more detailed
discussion of the method and how we used it is given elsewhere in this
volume by Toddy et al. (2012).

\begin{figure}
\centering
\includegraphics[scale=0.27]{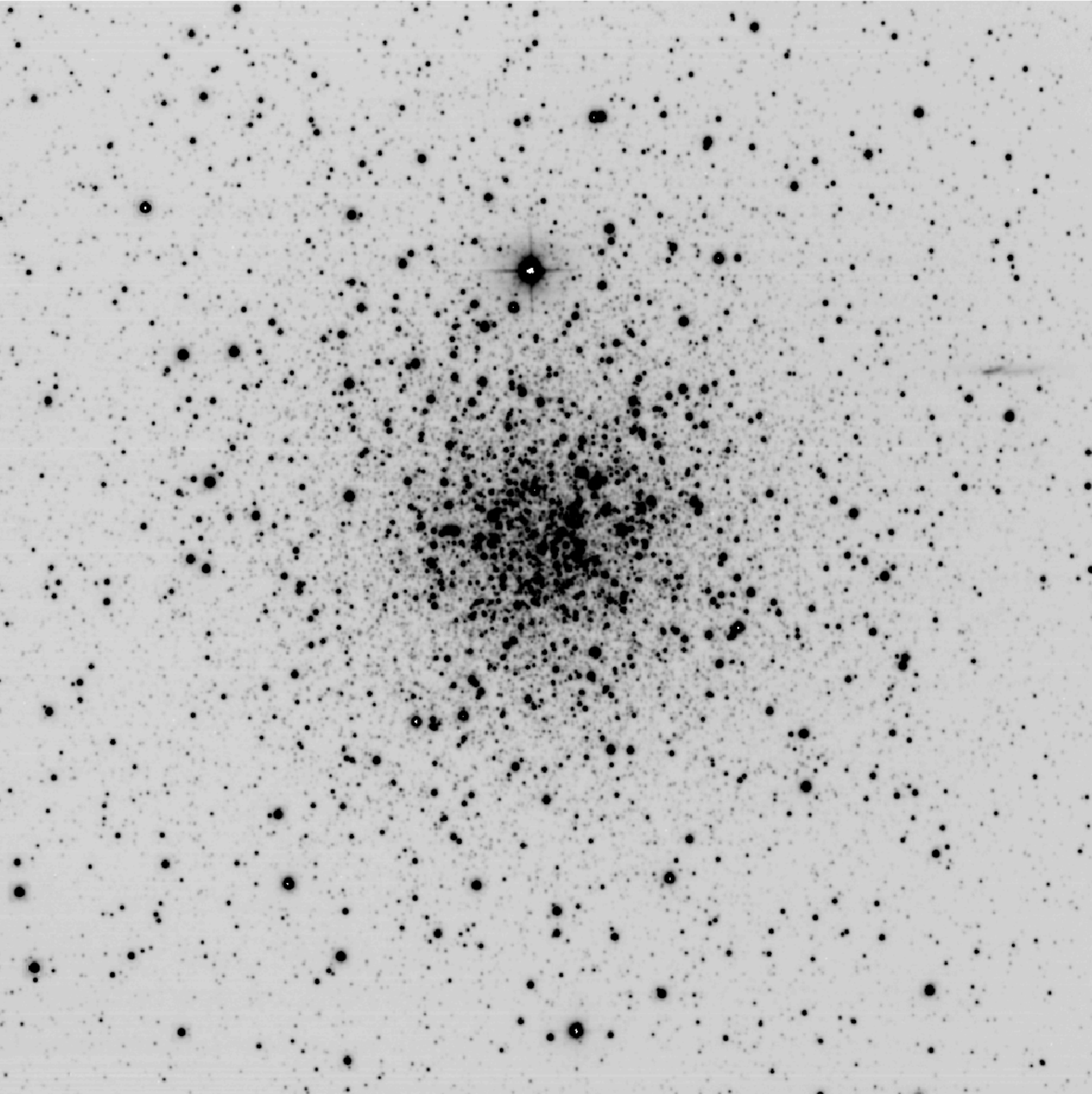}
\caption{The reference image used for all nights of the analysis/image
  subtraction.  This image is a combination of the best seeing images,
  with the combined seeing near $1.3\arcsec$. Note that brightest
  stars in the image are overexposed.}
\label{reffits}
\end{figure}

Due to the more than 4-month observation span, we were able to
determine the periods of most of the variables with periods less than
1 day to an accuracy of $10^{-5}$ days.  Those variables exhibiting
the Bla{\v z}hko Effect were less accurate due to modulation of their
light curves (Bla{\v z}hko 1907, Smith 2004).  Demers \& Wehlau's (1977) variables were identified
visually and astrometrically and then confirmed by comparing their
periods with our own.  Several of the variables identified by Demers
\& Wehlau (1977) were outside the field of view of our images thus
were not be observed. These include their variables V1, V2, V8 and V10.
We were not able to produce time-series photometry of two other
variables, V9 and V16.  Though they were in our field of view they
were overexposed giants.

\section{Results}

\begin{figure}
\centering
\includegraphics[scale=0.27]{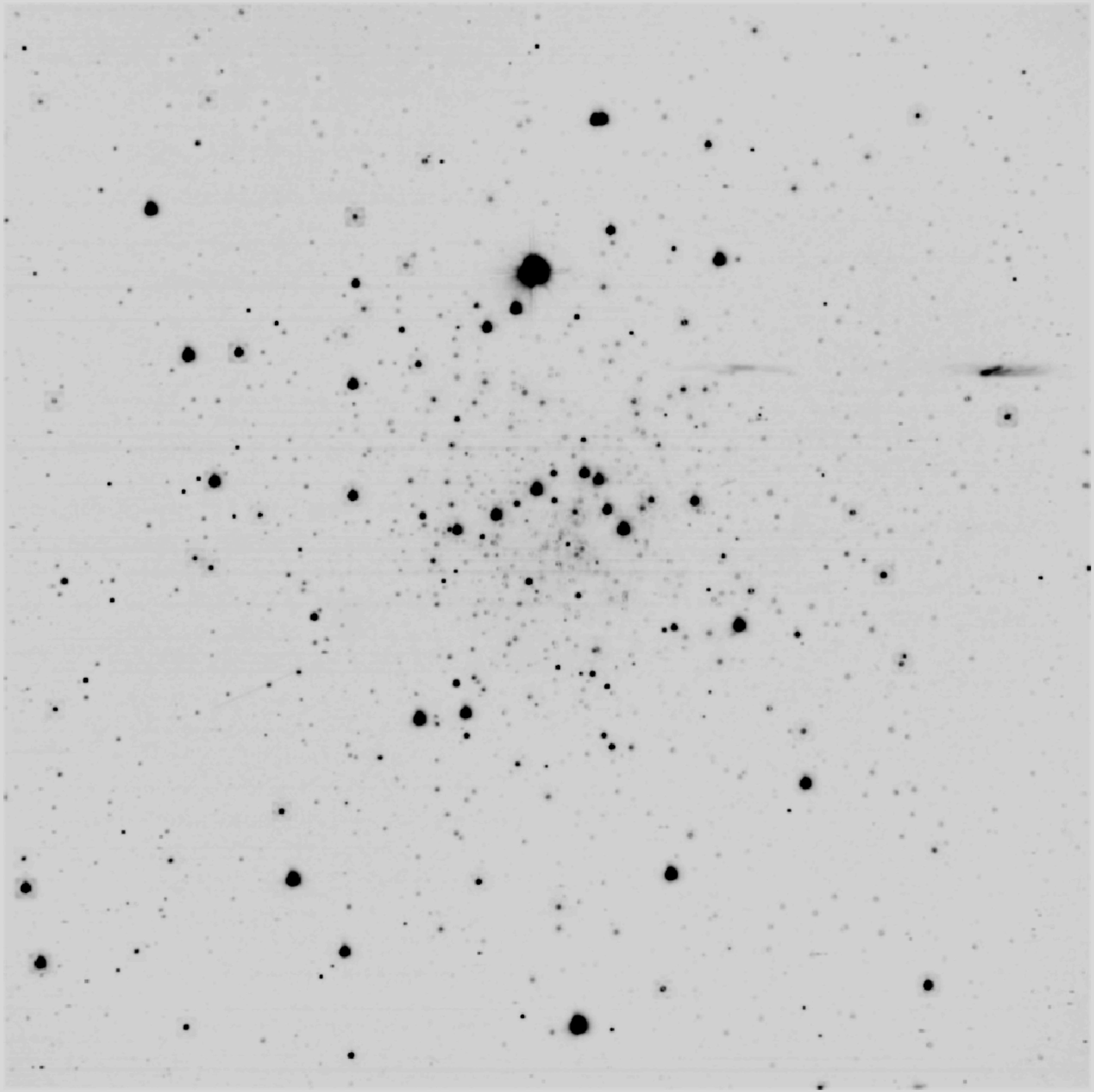}
\caption{The {\tt var.fits} frame output from ISIS used for image 
  subtraction of all 14 nights
  from SARA CTIO.  The relative amount of variation is indicated by
  the brightness of the star.  The overexposed stars show significant
  variation only because ISIS is unable to fit an accurate point
  spread function to them.}
\label{varfits}
\end{figure}

\begin{deluxetable}{lrc}
\tablecaption{Classification Summary}
\tablewidth{0pc}
\tablehead{
\colhead{Variable Type} & \colhead{Count} & \colhead{Period (days)} 
}
\startdata
RRab       & 10 & 0.69591\\
RRc          &   7 & 0.39555\\
RRe          &   2 & 0.30950\\
Eclipsing  &   2 & 0.32431\\
SX Phe     &   6 & 0.05847
\enddata
\end{deluxetable}

\begin{figure*}
\includegraphics[scale=0.83]{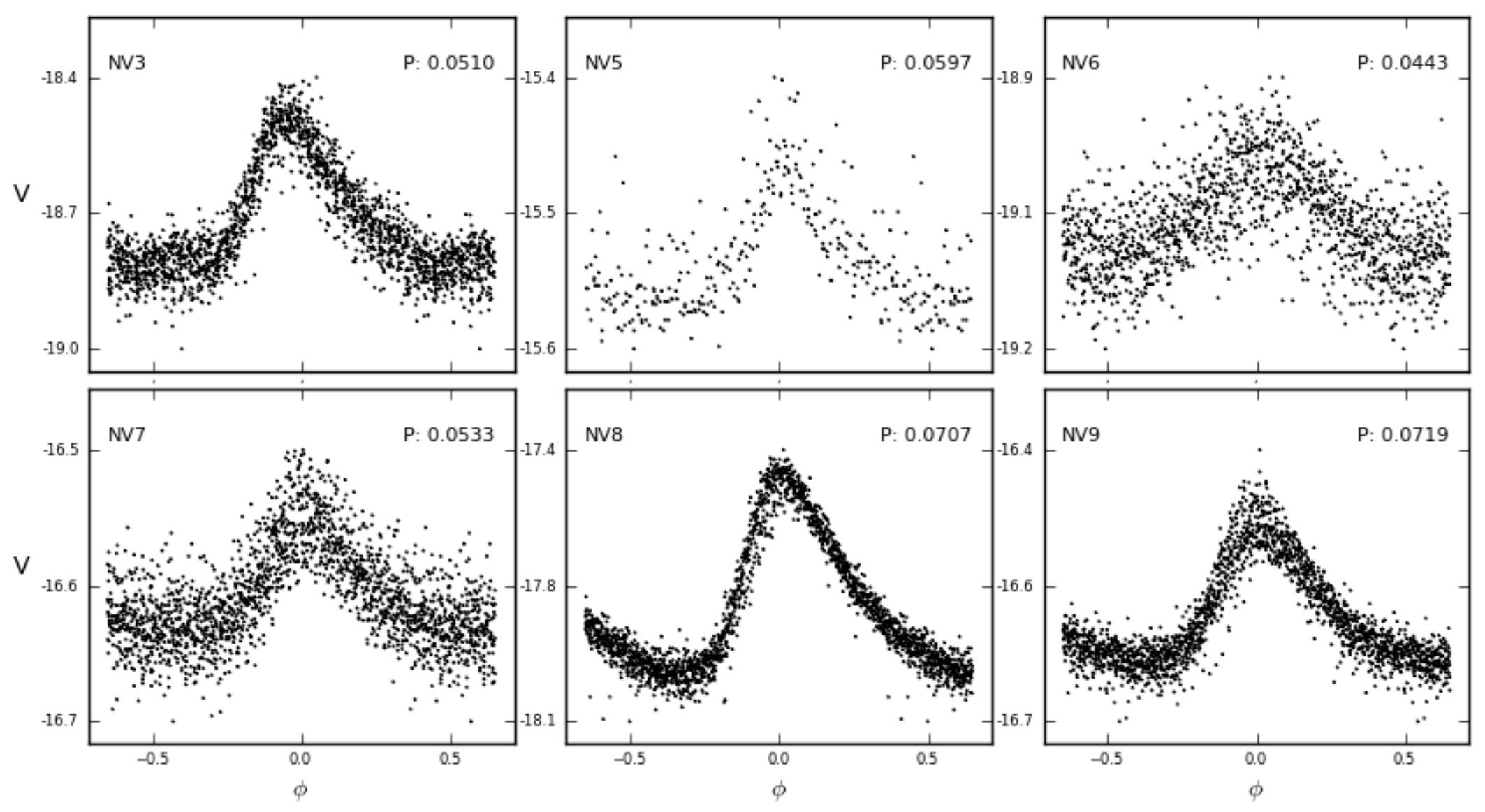}
\caption{6 previously unknown SX Phoenicis stars were found. Their mean period is 0.05847 days.}
\label{sx_phe}
\end{figure*}

\begin{figure*}

\includegraphics[scale=0.83]{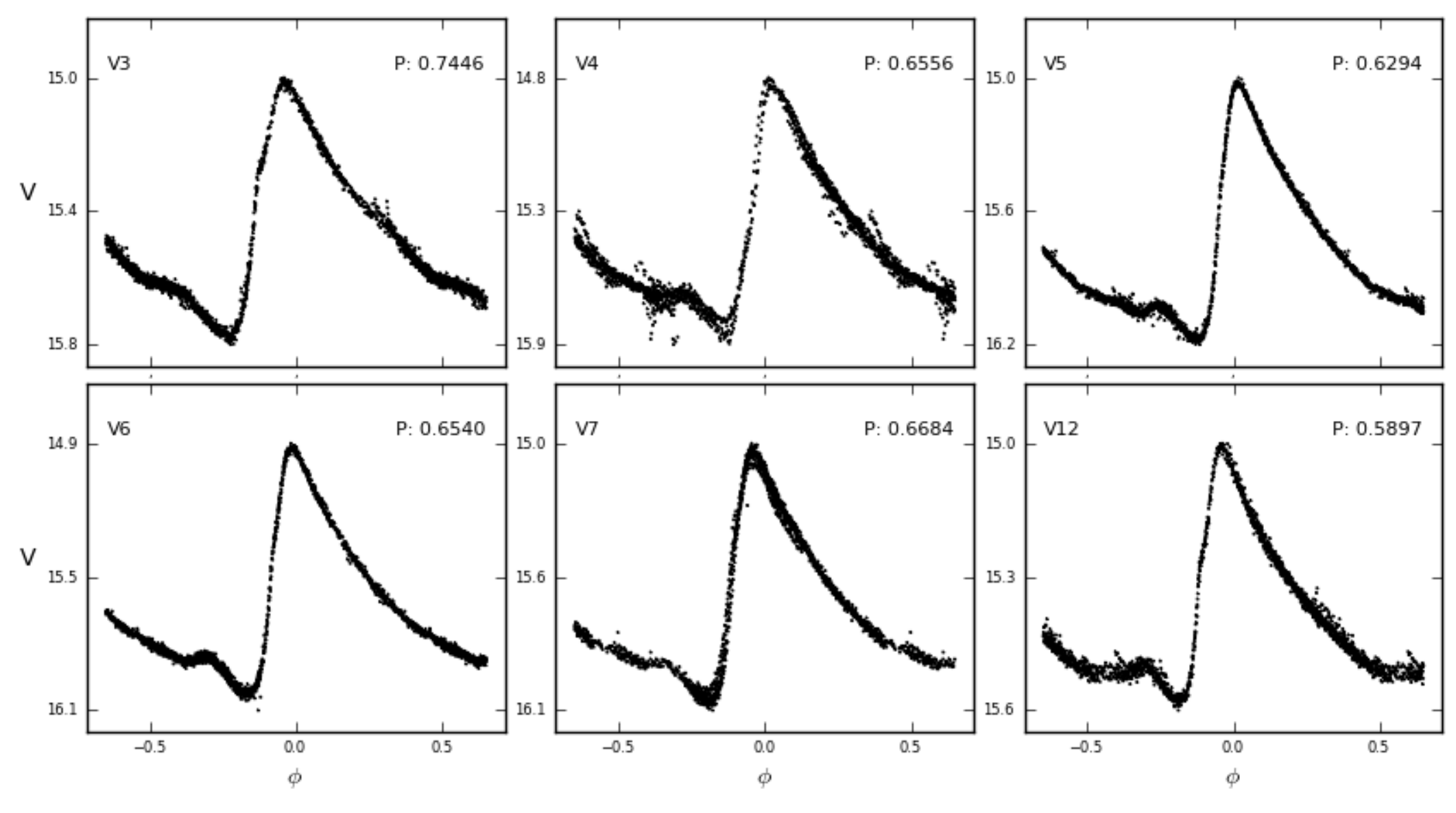}
\caption{Phased light curves for other variables
  with periods under a day.}
\label{phased_variables}
\end{figure*}

\begin{figure*}
\setcounter{figure}{3}
\includegraphics[scale=0.83]{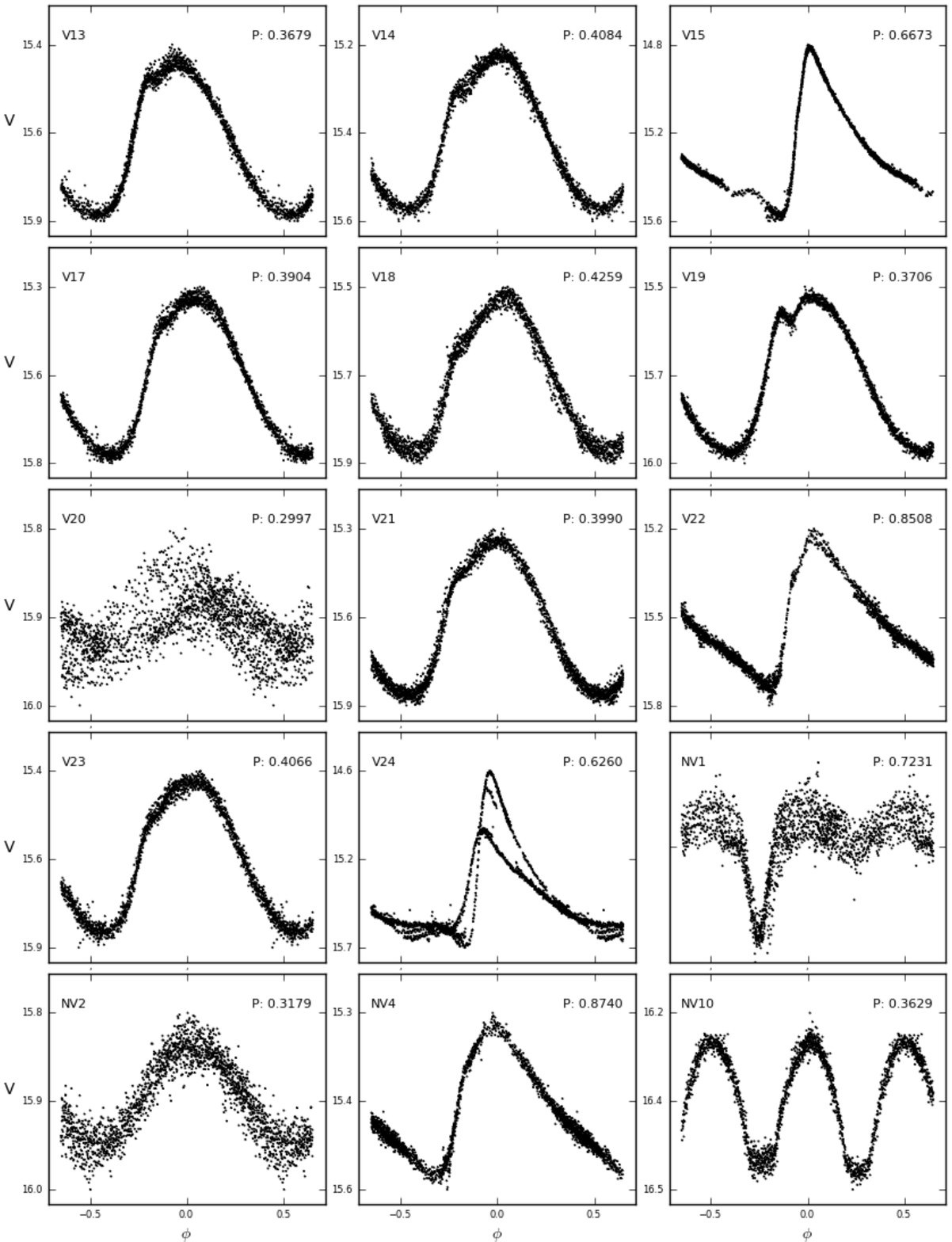}
\caption{(Continued) Phased light curves of all other variables with periods under 1 day.  Newly discovered variables have been given the NV prefix.}
\label{phased_variables}
\end{figure*}
Table 1 summarizes the results of our study.  In total we detected 27
variables with 10 of them being newly identified.  Of the variables
that were in our field of view and listed by Demers \& Welhau (1977)
we were able to identify all but two.  The two that we did not detect happened 
to be overexposed.  Of the total we detected 6 were newly discovered SX Phoenicis 
variables, 19 RR Lyrae variables, and 2 eclipsing variables.    
Table 2 gives the properties of each of the variables.
The SX Phoenicis variables are characterized by
short periods and relatively low amplitudes. The mean period of the SX
Phoenicis variables is 0.05847 days, with a minimum of 0.04425 days
and a maximum of 0.07191 days along with V amplitudes ranging from 0.2
to 0.7 magnitudes. These amplitudes are only estimates given that
many of the SX Phoenicis variables appear to be modulated and 
blended with other stars. NV5 was 
detected by ISIS only on 5 April.  The others were detected on all of the nights.  All of
the SX Phoenicis stars were within the half-mass radius (2.4{\arcmin}) of the cluster as
listed by Harris (1996).   The phased light curves for the  SX 
Phoenicis stars are shown in Figure 3. 

\begin{figure}
\centering
{\includegraphics[scale=0.48]{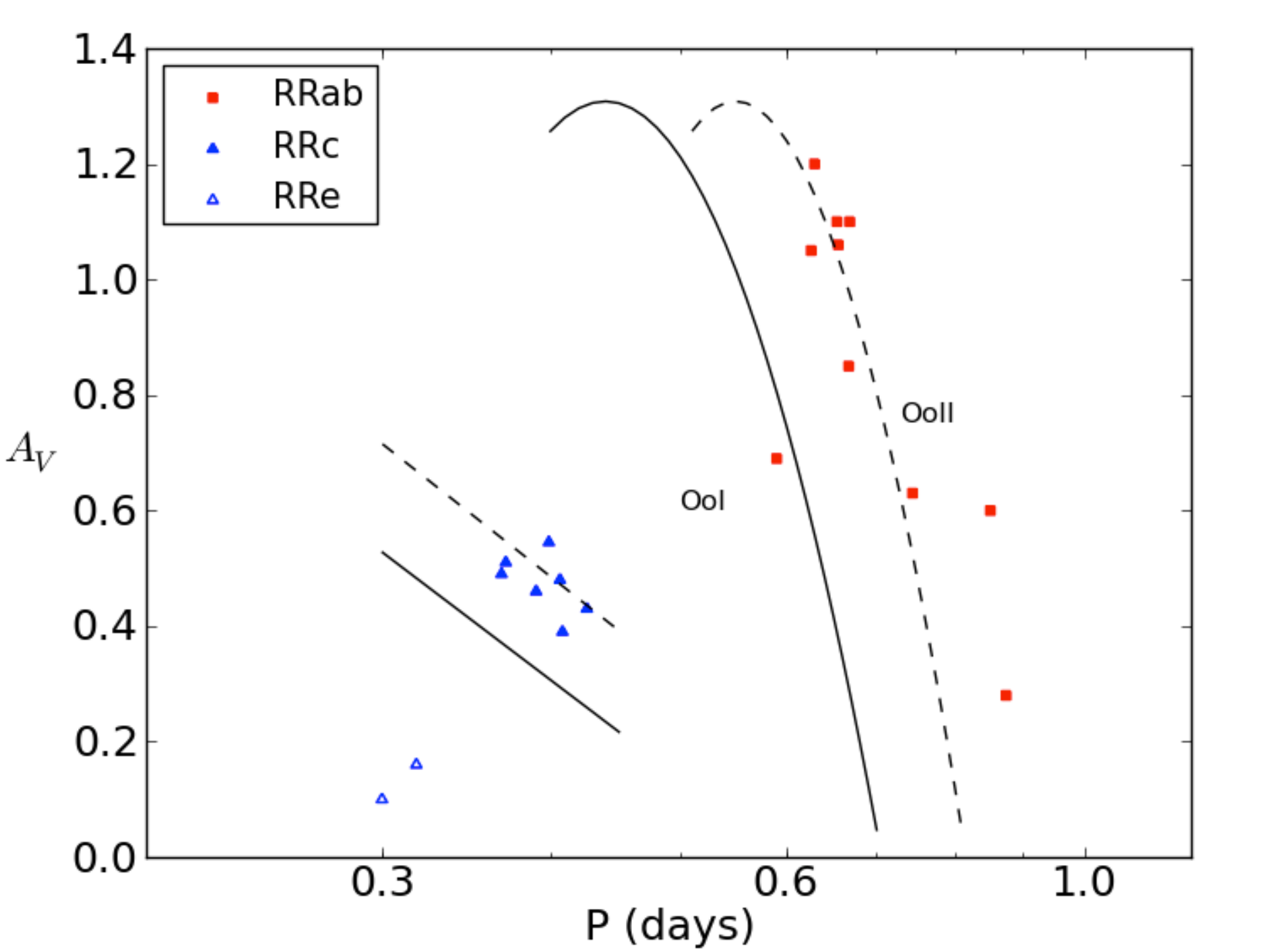}}
\caption{Bailey diagram for the RR Lyrae variables in NGC~4833.
  Oosterhoff I and II lines are from fitting formula of Zorotovic et
  al. (2010). RRab and RRc period-amplitude relations are indicative
  of NGC 4833 being an Oosterhoff type II cluster.  Note that our two
  suspected RRe variables have much lower amplitudes and slightly
  shorter periods than the RRc variables.}
\label{lpvs}
\end{figure}

\begin{figure}
\centering
\includegraphics[scale=0.49]{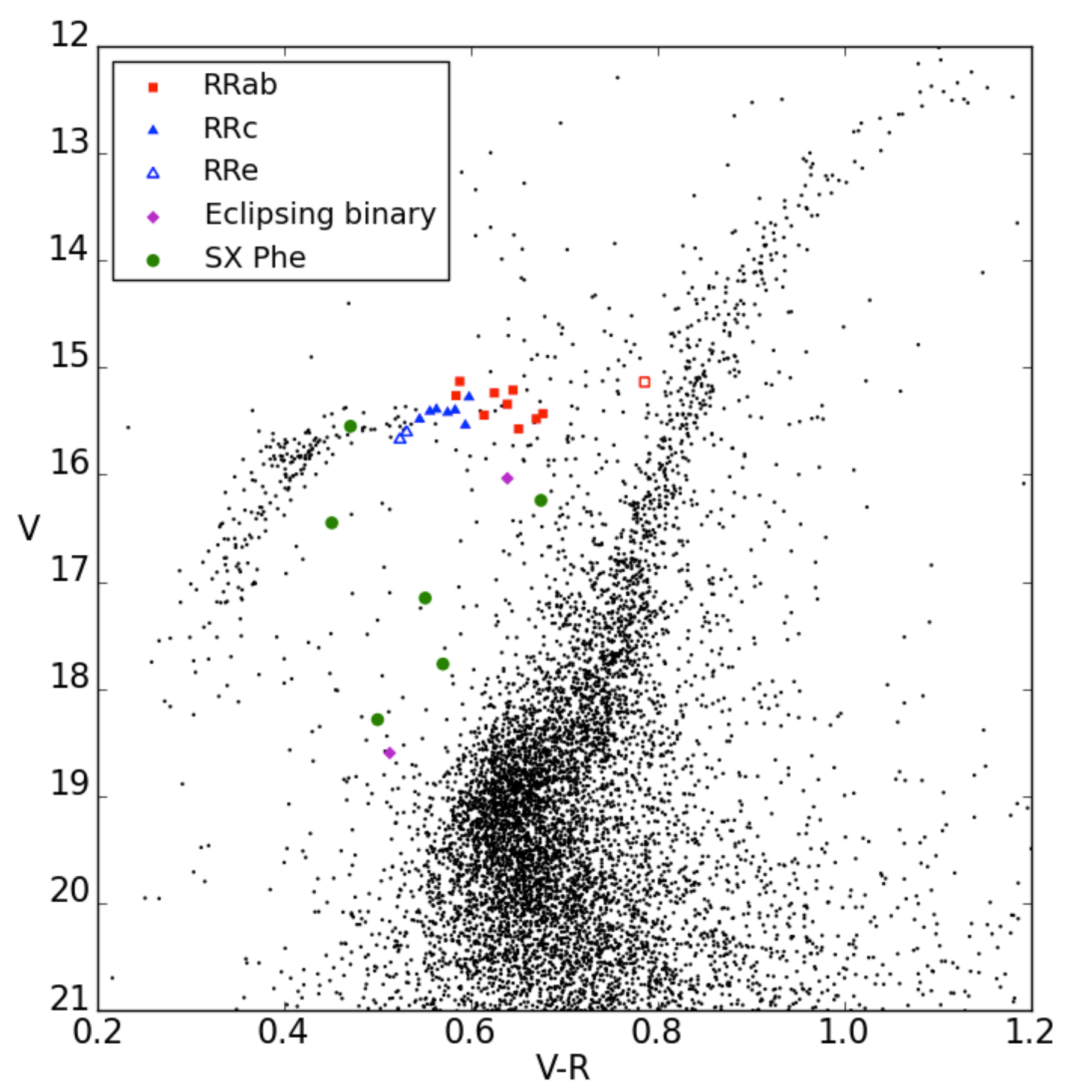}
\caption{Color-magnitude diagram for NGC~4833, with variable stars
  indicated. 165 V band images, with an exposure 120 seconds, and 175
  R band images, with an exposure of 45 seconds were combined together
  to produce this CMD. Appreciable differential reddening across the
  cluster, as was noted by Melbourne et al. (2000), causes significant
  variation of the position of variables along the horizontal branch.}
\label{period_hist}
\end{figure}

Of the RR Lyrae variables identified in the cluster 10 were RRab 
variables, with a mean period of 0.69591 days,
7 were RRc variables, with a mean period of 0.39555 days, and variable V20
which may be a second overtone (RRe) variable previously noted by Demers \&
Wehlau (1977).  But given V20's modulated light curve it may be a type RRd double-mode 
variable (Alcock et al. 1996, Bono et al. 1997).  Our newly discovered variable stars include 
a W Ursae Majoris star (period of 0.3629 days), an RRab star (period
of 0.8740 days), another second overtone (RRe) variable (period of 0.3179
days), and an Algol type binary star (period of 0.7231 days). In total we identified 10
new variables in the cluster.   The phased light curves for these variables are shown in Figure 4.  
In addition to the  ratio of RRab to RRc Lyrae 
variables the Bailey diagram presented in Figure 5 shows the cluster is of 
the Oosterhoff type II cluster.    In this diagram the fitting formulae are from 
Zorotovic et al. (2010). Figure 5 also shows the two outlying RRe variables, 
indicated by the open triangles.

Color-magnitude diagrams (CMD) are useful in determining cluster membership.  
To verify cluster membership we produced a Color-magnitude diagram for
NGC 4833 shown in Figure 6 with the various types of variable stars indicated. To construct
this diagram we combined 165 V band images, with an exposure of 120
seconds, and 175 R band images, with an exposure of 45 seconds. The
images were obtained as part of a BVRI color sequence on 9 March, 1,
3, and 6 May, 2011.  {\tt DAOPHOT} was used to determine V and R
magnitudes of the stars (Stetson 1987).  In the CMD shown in Figure 6
RRab and RRc variables are indicated by the red and blue colored
points, respectively.  RRc variables tend to lie on the warmer (left)
side of the instability strip whereas RRab variables are to the cooler
(right) of the instability strip. Scatter in the variables is partly
due to blending of the variables with other stars, our averaging of
several hundred images before using {\tt DAOPHOT}, and possible
variable reddening in the direction of the cluster. 
We assumed fixed atmospheric extinction coefficients for each color and for all nights. 

\setlongtables
\begin{longtable}{cccccc}
\tabletypesize{\scriptsize}
\tablecaption{Variables Stars in NGC 4833}
\tablewidth{0pt}
\tablehead{
\colhead{V$\#$}& 
\colhead{RA(h,m,s)} & 
\colhead{Dec($^{\circ}$,$\arcmin$,$\arcsec$)} &
\colhead{P(d)} & 
\colhead{V~~~A$_V$} &
\colhead{Type} 
}
\startdata
V1 & \multicolumn{5}{l}{RY Musca, not in our field of view.}\\
V2 & \multicolumn{5}{l}{RZ Musca, not in our field of view.}\\
V3   & 12 59 33.68 & -70 52 13.8 & 0.74449 & 15.1  0.63 & RRab \\
V4   & 12 59 33.72 & -70 51 58.4 & 0.65550 & 15.3  1.06 & RRab \\
V5   & 13 00 01.26 & -70 53 17.7 & 0.62942 & 15.3  1.20 & RRab \\
V6   & 12 59 56.12 & -70 50 10.5 & 0.65395 & 15.6  1.10 & RRab \\
V7   & 12 59 48.79 & -70 52 21.5 & 0.66842 & 15.4  1.10 & RRab \\
V8   & \multicolumn{5}{l}{Confirmed not to be variable.}\\
V9  &  \multicolumn{5}{l}{Overexposed long period variable.}\\
V10 & \multicolumn{5}{l}{Not in our field of view.}\\
V11 & \multicolumn{5}{l}{Not in our field of view.}\\
V12 & 12 59 37.99 & -70 52 15.1 & 0.58970 & 15.4  0.69 & RRab \\
V13  & 13 00 29.79 & -70 52 56.1 & 0.36788 & 15.4  0.49 & RRc  \\
V14  & 12 59 31.02 & -70 53 07.3 & 0.40841 & 15.3  0.39 & RRc  \\
V15  & 12 59 20.10 & -70 53 25.6 & 0.66730 & 15.1  0.85 & RRab \\
V16 & \multicolumn{5}{l}{Overexposed long period variable.}\\
V17  & 12 59 43.96 & -70 54 25.5 & 0.39039 & 15.4  0.46 & RRc  \\
V18  & 12 59 28.19 & -70 54 26.2 & 0.42587 & 15.5  0.43 & RRc  \\
V19  & 12 59 05.94 & -70 53 30.3 & 0.37060 & 15.5  0.51 & RRc  \\
V20  & 12 59 08.10 & -70 52 24.2 & 0.30020 & 15.7  0.10 & RRe  \\
V21  & 12 59 50.90 & -70 50 36.8 & 0.39900 & 15.4  0.55 & RRc  \\ 
V22  & 12 59 45.08 & -70 53 55.9 & 0.85070 & 15.2 0.60 & RRab\\
V23  & 12 59 44.72 & -70 51 27.5 & 0.40665 & 15.4 0.48 & RRc\\ 
V24  & 12 59 36.67 & -70 52 59.1 & 0.62570 & 15.5 1.05 & RRab\\
NV1  & 12 58 55.31 & -70 51 46.4 & 0.72300 & 18.6 0.00 & EA\\
NV2  & 12 59 02.68 & -70 52 52.3 & 0.31790 & 15.6 0.16 & RRe\\
NV3  & 12 59 13.65 & -70 52 10.3 & 0.05097 & 17.8 0.62 & SXPhe\\
NV4  & 12 59 21.20 & -70 53 26.8 & 0.87395 & 15.2 0.28 & RRab\\
NV5  & 12 59 35.32 & -70 52 41.9 & 0.05965 & 15.5 0.10 & SXPhe\\
NV6  & 12 59 42.52 & -70 53 04.6 & 0.04425 & 18.3 0.30 & SXPhe\\
NV7  & 12 59 47.92 & -70 52 52.2 & 0.05332 & 16.5 0.19 & SXPhe\\
NV8  & 12 59 55.04 & -70 52 24.6 & 0.07067 & 17.2 0.70 & SXPhe\\
NV9  & 12 59 57.66 & -70 54 30.6 & 0.07191 & 16.2 0.33 & SXPhe\\
NV10 & 13 00 25.08 & -70 49 16.3 & 0.36287 & 16.0 0.15 & EC
\end{longtable}
\enddata


Of the RR Lyrae variables only one, V15, appears to be a blend.  V15 is
shown in Figure 6 as the hollow red square to the right of the
instability strip.  Other than this errant variable the remainder
appear near the instability strip.  In addition to blending there is
significant extinction across the cluster.  The E(B-V) is listed as
0.32 and Melbourne et al.  (2000) found that this varies depending on
location.  Because of this variable interstellar extinction some scatter is introduced into our mean
magnitudes of the RR Lyrae variables. 

The SX Phoenicis variables all lie to the left of the giant branch.
After close inspection of the images all but one of our SX Phoenicis
variables appeared to be blended with stars of equal or greater brightness.
Further inspection of Hubble Space
Telescope images confirmed this.  Variables NV4 and NV5 appear to be
blended with giant stars whereas NV7, NV8, and NV9 are blended with
stars of similar brightness to themselves.  NV6 is the only SX Phoenicis 
variable that does not appear to be blended with a relatively bright or 
comparable bright star.

\vspace{1cm}

\section{Conclusions}

For nearly 5 months from late January through June of 2011 we
observed the globular cluster NGC~4833 using the SARA 0.6
meter telescope at CTIO. Using the image subtraction software ISIS we
were able to find precise periods of 17 previously known variables and 10
newly discovered variables stars.  Of these new discoveries we found 6 SX 
Phoenicis variables.   The other 4 newly discovered variables were 2 eclipsing, one RRab, 
and one apparent RRe.  In total, we
classified 10 of the variables as type RRab, with a mean period of
0.69591 days, 7 as type RRc with a mean period of 0.39555 days,
perhaps 2 lower amplitude type RRe, with a mean period of 0.30950 days, and 2
eclipsing binaries, with a mean period of 0.3243 days.  Both the Bailey diagram and ratio of
$N_{c} /N_{ab}$ are consistent with the cluster being of type
Oosterhoff Type II.




\acknowledgments

We thank C.  Alard for making ISIS 2.2 publicly available.  This
project was funded in part by the Butler Institute for Research and
Scholarship.  The authors also thank F.  Levinson for a generous gift
enabling Butler University's membership in the SARA consortium.





\end{document}